\documentclass[a4paper,11pt]{article}
\usepackage{amssymb}
\usepackage{graphicx}
\usepackage{amsmath}

 \DeclareFontFamily{U}{rsf}{}
\DeclareFontShape{U}{rsf}{m}{n}{
  <5> <6> rsfs5 <7> <8> <9> rsfs7 <10-> rsfs10}{}
\DeclareMathAlphabet\Scr{U}{rsf}{m}{n} \makeatletter
\@addtoreset{equation}{section} \makeatother

\def\be{\begin{equation}}
\def\ee{\end{equation}}
\def\ba{\begin{array}}
\def\ea{\end{array}}

\newcommand{\bea}{\begin{eqnarray}}
\newcommand{\eea}{\end{eqnarray}}

\begin{document}

\begin{titlepage}
 \thispagestyle{empty}
\begin{flushright}
     \hfill{CERN-PH-TH/2011-216}\\
 \end{flushright}

 \vspace{100pt}

 \begin{center}
     { \huge{\bf      {Generalized Mirror Symmetry\\\vspace{6pt}and\\\vspace{10pt}Quantum Black Hole Entropy }}}

     \vspace{25pt}

     {\Large {Sergio Ferrara$^{a,b,c}$ and Alessio Marrani$^{a}$}}

     \vspace{40pt}

  {\it ${}^a$ Physics Department, Theory Unit, CERN,\\
     CH -1211, Geneva 23, Switzerland;\\
     \texttt{sergio.ferrara@cern.ch}\\
     \texttt{Alessio.Marrani@cern.ch}}

     \vspace{10pt}

    {\it ${}^b$ INFN - Laboratori Nazionali di Frascati,\\
     Via Enrico Fermi 40, I-00044 Frascati, Italy}

     \vspace{10pt}

     {\it ${}^c$  Department of Physics and Astronomy,\\
University of California, Los Angeles, CA 90095-1547,USA}\\


     \vspace{30pt}

     \vspace{60pt}

     {ABSTRACT}

 \vspace{10pt}
 \end{center}
We find general relations between the on-shell gravitational trace anomaly $A_{\mathcal{N}%
}$, and the logarithmic correction $\Delta S_{\mathcal{N}}$ to the
entropy of ``large" BPS extremal black holes in
$\mathcal{N}\geqslant 2$ supergravity theories in $D=4$
space-time dimensions (recently computed by Sen \cite{Sen-2}). For (generalized) self-mirror theories (all having $%
A_{\mathcal{N}}=0$), we obtain the result $\Delta
S_{\mathcal{N}}=-\Delta S_{8-\mathcal{N}}=2-\mathcal{N}/2$, whereas
for generic theories the trace anomaly $\widetilde{A}_{\mathcal{N}}$
of the fully dualized theory turns out
to coincide with $2\Delta S_{\mathcal{N}}$, up to a model-independent shift: $%
\widetilde{A}_{\mathcal{N}}=2\Delta S_{\mathcal{N}}-1$. We also
speculate on $\mathcal{N}=1$ theories displaying ``large" extremal
black hole solutions.
\end{titlepage}

\section{\label{Intro}Introduction}

Recently, a generalized notion of mirror symmetry was suggested \cite{DF-1},
under which
\begin{equation}
A_{\mathcal{N}}=-\frac{1}{24}\rho ,
\end{equation}
occurring in the on-shell\footnote{%
As given by Eq. (\ref{def-A}), we call ``on-shell'' anomaly the one
concerning the square of $R_{\mu \nu \rho \sigma }$, following \cite{DF-1}
(see also \textit{e.g.} \cite{Duff-2} for an extensive list of Refs.).
\par
It should be pointed out that this is not the same as the anomaly computed
on the supergravity equations of motion. Indeed, while the coefficient of
the Gauss-Bonnet term is always proportional to $n_{s}+62n_{V}+\frac{11}{2}%
n_{MF}$ ($n_{s}$, $n_{V}$ and $n_{MF}$ respectively standing for the number
of scalar, vector and Majorana spinor massless fields), in a conformally
flat background (as is the Bertotti-Robinson $AdS_{2}\times S^{2}$
near-horizon geometry of the extremal black hole), the term proportional to
the square of the Weyl tensor does vanish (see \textit{e.g.} \cite{Duff-2},
and also \cite{Sen-2} for a recent discussion).} gravitational trace anomaly
\cite{Duff-1,Duff-2,Duff-3}
\begin{equation}
g_{\mu \nu }\left\langle T^{\mu \nu }\right\rangle =A_{\mathcal{N}}\frac{1}{%
32\pi ^{2}}R^{\ast \mu \nu \rho \sigma }R_{\mu \nu \rho \sigma }^{\ast },
\label{def-A}
\end{equation}
changes sign.

In $M$-theory compactified on a seven-manifold $X^{7}$ with Betti numbers $%
\left( b_{0},b_{1},b_{2},b_{3}\right) $, $\rho $ is defined as \cite{DF-1}
\begin{equation}
\rho \equiv 7b_{0}-5b_{1}+3b_{2}-b_{3},
\end{equation}
and $\rho \rightarrow -\rho $ under the generalized mirror symmetry \cite
{DF-1}
\begin{equation}
\left( b_{0},b_{1},b_{2},b_{3}\right) \rightarrow \left(
b_{0},b_{1},b_{2}-\rho /2,b_{3}+\rho /2\right) .
\end{equation}

In Ref. \cite{DF-1} it was shown that $D=4$, $\mathcal{N}\geqslant 3$
extended supergravity theories are generalized self-mirror\footnote{%
For $\mathcal{N}=3$, $4$, this is true provided at least $n_{V}=2$, $3$
vector multiplets are present in the (fully) dualized theories.}. On the
other hand, for $\mathcal{N}=1$, $2$ theories the generalized self mirror
condition imposes some constraints on the matter content.

As we will see below, results crucially depend on the dualization of $3$-
and $2$- form fields, which naturally arise from $M$-theory
compactifications; it is remarkable that the trace anomaly coefficient $A_{%
\mathcal{N}}$ of the undualized theory does vanish for $\mathcal{N}=8$, $6$
and $5$ ``pure'' supergravities, if the corresponding graviton multiplet is
properly defined as containing also form fields of degree higher than one
\cite{DVN} (see also \cite{DF-1}). This is still true in matter coupled $%
\mathcal{N}=3$ and $4$ theories, if at least $n_{V}=2$ resp. $3$ massless
vector multiplets in the dualized theory (corresponding to $2$ resp. $3$
massless $2$-form multiplets in the undualized \textit{avatar}) are taken
into account. Self-mirror $\mathcal{N}=2$ theories and generalized
self-mirror $\mathcal{N}=1$ theories all have $\rho =0$, which in the fully
dualized framework respectively contrains the matter content as follows \cite
{DF-1}:
\begin{eqnarray}
\mathcal{N} &=&2:n_{H}=n_{V}+1; \\
\mathcal{N} &=&1:n_{c}=3n_{V}+7,
\end{eqnarray}
where $n_{V}$, $n_{c}$ and $n_{H}$ respectively denote the number of vector,
chiral and hyper massless multiplets.\medskip

On the other hand, Sen \textit{et al.} \cite{Sen-0,Sen-1,Sen-2} recently
computed the coefficient $\Delta S$ of the logarithmic correction to the
Bekenstein-Hawking \cite{BH} entropy of ``large'' BPS extremal black holes
(BHs), in particular achieving the following result for a generic $\mathcal{N%
}=2$ supergravity:
\begin{equation}
\Delta S_{\mathcal{N}=2}=\frac{1}{24}\left( 23+n_{H}-n_{V}\right) .
\label{Sen-N=2}
\end{equation}
For self-mirror $\mathcal{N}=2$ theories ($n_{H}=n_{V}+1$), Eq. (\ref
{Sen-N=2}) yields
\begin{equation}
\text{self-mirror}:\Delta S_{\mathcal{N}=2}=1,  \label{N=2-self-mirror}
\end{equation}
as it holds for the self-mirror $stu$ model \cite{Sen-1}, characterized by $%
n_{V}=3$ and $n_{H}=4$.

Up to some irrelevant $\mathcal{O}\left( 1\right) $ terms, the following
structure is known to hold in general (see \textit{e.g.} \cite
{Sen-0,Sen-1,Sen-2}, and Refs. therein):
\begin{equation}
S=S_{0}+\Delta S\ln \left( A_{H}^{2}\right) ,  \label{gen-1}
\end{equation}
where $S_{0}=A_{H}/4$ is the Bekenstein-Hawking entropy of the ``large'' BPS
extremal BH under consideration, whose non-vanishing event horizon area is
denoted by $A_{H}$. Due to the Attractor Mechanism \cite{AM-Refs,FGK}, $%
A_{H} $ is given by the BH effective potential $V_{BH}$ computed at its
attractor points \cite{FGK}: $A_{H}=4\pi \left. V_{BH}\right| _{\partial
V_{BH}=0}$. Actually, in any theory with $\mathcal{N}\geqslant 3$ the scalar
manifold is a symmetric coset $\frac{G_{4,\mathcal{N}}}{H_{4,\mathcal{N}}}$,
and it holds that $A_{H}=\sqrt{I_{4}}$, where $I_{4}$ is the unique
independent polynomial invariant (quartic in electric and magnetic charges)
constructed with the BH charge irrepr. of $G_{4,\mathcal{N}}$. The symmetric
coset structure of the scalar manifold also characterizes $\mathcal{N}=2$
minimally coupled and $\mathcal{N}=3$ matter coupled theories, but in such
theories $\sqrt{I_{4}}=\left| I_{2}\right| $. In general, the scalar
manifold of $\mathcal{N}=2$ and $\mathcal{N}=1$ theories, despite being
characterized by K\"{a}hler geometry (of special type in $\mathcal{N}=2$),
is not necessarily symmetric nor homogeneous, and $I_{4}$ may thus not exist
at all.\medskip

Aim of the present note is to clarify the relation between $A_{\mathcal{N}}$
and $\Delta S_{\mathcal{N}}$ for $\mathcal{N}\geqslant 2$, and consider,
within some consistency conditions, also $\mathcal{N}=1$ theories of gravity
exhibiting ``large'' extremal BH solutions.

Two main general results are achieved in this investigation:

\textbf{I]} (Generalized) self-mirror theories exhibit $\mathcal{N}$%
-dependent values of $\Delta S_{\mathcal{N}}$ related by a fermionic
symmetry:
\begin{equation}
\text{(gen.)~self-mirror}:\Delta S_{\mathcal{N}}=-\Delta S_{8-\mathcal{N}}=2-%
\frac{\mathcal{N}}{2}.  \label{main-1}
\end{equation}
this result can be made explicit by the following symmetric pattern,
centered at $\mathcal{N}=4$:
\begin{equation}
\begin{array}{cccccccccc}
\mathcal{N}: & 8 & \left( 7\right) & 6 & 5 & 4 & 3 & 2 & 1 & 0 \\
\Delta S_{\mathcal{N}}: & -2 & \left( -\frac{3}{2}\right) & -1 & -\frac{1}{2}
& 0 & \frac{1}{2} & 1 & \frac{3}{2} & 2?
\end{array}
,  \label{DeltaS-N-scheme}
\end{equation}
suggesting a possible ``generalized self-mirror'' $\mathcal{N}=0$, $D=4$
gravity theory with $\Delta S_{\mathcal{N}=0}=2$.

\textbf{II]} In generic theories, the gravitational trace anomaly $%
\widetilde{A}_{\mathcal{N}}$ of the fully dualized theory is nothing but $%
\Delta S_{\mathcal{N}}$ itself, up to a model-independent shift:
\begin{equation}
\widetilde{A}_{\mathcal{N}}=2\Delta S_{\mathcal{N}}-1.  \label{main-2}
\end{equation}
Note that $\widetilde{A}_{\mathcal{N}}$ is the value of the on-shell
gravitational trace anomaly coefficient as computed in ``standard'' $D=4$
supergravity theories, with only physical spin degrees of freedom (see App.
A of \cite{DF-1} for a detailed treatment). In $M$-theory on $X^{7}$, the
degrees of freedom $f$, the number $\#$ and the contribution to $\widetilde{A%
}_{\mathcal{N}}$ of the various types of massless fields in the fully
dualized $\mathcal{N}\geqslant 1$, $D=4$ supergravity theory are given in
Table XX of \cite{DF-1}, which we partially report in Table 1.

\begin{table}[h]
\begin{center}
\begin{tabular}{|c||c|c|c|}
\hline
$
\begin{array}{ccc}
&  &
\end{array}
$ & $
\begin{array}{ccc}
& f &
\end{array}
$ & $
\begin{array}{ccc}
& 360\widetilde{A}_{\mathcal{N}} &
\end{array}
$ & $
\begin{array}{ccc}
& \# &
\end{array}
$ \\ \hline\hline
$g_{\mu \nu }$ & $2$ & $848$ & $b_{0}$ \\ \hline
$A_{\mu }$ & $2$ & $-52$ & $b_{1}+b_{2}$ \\ \hline
$\phi $ & $1$ & $4$ & $2b_{3}$ \\ \hline
$\psi _{\mu }$ & $2$ & $-233$ & $b_{0}+b_{1}$ \\ \hline
$\chi $ & $2$ & $7$ & $b_{2}+b_{3}$ \\ \hline
\end{tabular}
\end{center}
\caption{Degrees of freedom $f$, contribution to $\widetilde{A}_{\mathcal{N}%
} $ and number $\#$ of the various massless fields in a fully dualized $%
\mathcal{N}\geqslant 1$, $D=4$ supergravity theory obtained as
compactification of $M$-theory on $X^{7}$ with Betti numbers $\left(
b_{0},b_{1},b_{2},b_{3}\right) $ \protect\cite{DF-1}. }
\end{table}
By virtue of (\ref{main-1}), for (generalized) self-mirror theories (\ref
{main-2}) can be recast as
\begin{equation}
\text{(gen.)~self-mirror}:\widetilde{A}_{\mathcal{N}}=3-\mathcal{N},
\label{main-2-self-mirror}
\end{equation}
thus curiously yielding $\widetilde{A}_{\mathcal{N}=3}=0$ (as noted long
time ago in \cite{Duff-3}). It is worth observing that $\widetilde{A}_{%
\mathcal{N}=8}=-5$ matches the result of \cite{Christensen-1}; in
particular, as given by the general formula (\ref{main-2}), $\widetilde{A}_{%
\mathcal{N}=8}$ is not proportional to $\Delta S_{\mathcal{N}=8}$.\medskip

The plan of this note is as follows.

In Sec. \ref{N=2-Multiplet-Decomp}, starting from some results recently
obtained in \cite{Sen-0,Sen-1,Sen-2}, the massless multiplet content of
fully dualized $\mathcal{N}\geqslant 3$, $D=4$ supergravity theories is
decomposed in terms of the various types of $\mathcal{N}=2$ multiplets,
whose contributions to $\Delta S_{\mathcal{N}=2}$ are then explicitly
computed.

General relations, involving $\Delta S_{\mathcal{N}}$, the undualized trace
anomaly $A_{\mathcal{N}}$ and the fully dualized trace anomaly $\widetilde{A}%
_{\mathcal{N}}$, are obtained in Sec. \ref{General-Rels}.

Two classes of $\mathcal{N}=1$ theories are treated in Sec. \ref
{N=1-Theories}. Subsec. \ref{N=1-as-Trunc-N=2} deals with consistent $%
\mathcal{N}=1$ truncations of $\mathcal{N}=2$ theories, and a general
formula for $\Delta S_{\mathcal{N}=1}$ is obtained; this is the class of $%
\mathcal{N}=1$ theories for which the general results derived in Sec. \ref
{General-Rels}, specified for $\mathcal{N}=1$, hold. Another class of $%
\mathcal{N}=1$ theories, which we dub ``minimally coupled'', is then
considered in Subsec. \ref{N=1-Theories-mc}, and the corresponding $\Delta
S_{\mathcal{N}=1}^{\text{mc}}$ is computed; by performing a proper $\mathcal{%
N}=0$ limit, the result $\Delta S_{\mathcal{N}=0}$, recently computed in
\cite{Sen-2}, is recovered.

\section{\label{N=2-Multiplet-Decomp}$\mathcal{N}=2$ Multiplet Decomposition
of $\mathcal{N}\geqslant 3$ Theories}

A crucial step in the treatment given in \cite{Sen-2} is the fact that \cite
{Sen-0,Sen-1}
\begin{equation}
\Delta S_{\mathcal{N}=4}=0  \label{N=4-res}
\end{equation}
\textit{for any} number $n$ of coupled matter (vector) massless multiplets.

The various types of $\mathcal{N}=2$ massless multiplets will be referred to
as $G_{\lambda _{\max }}$, where $\lambda _{\max }$ denotes the maximal
helicity of the multiplet ($G\equiv G_{2}$, $G_{3/2}$, $G_{V}\equiv G_{1}$
and $G_{H}\equiv G_{1/2}$ respectively stand for the graviton, gravitino,
vector and hyper multiplets), in Table 2 the multiplet content of any fully
dualized $\mathcal{N}\geqslant 3$, $D=4$ supergravity theory in terms of
these building blocks is given (see \textit{e.g.} \cite
{N=8-red,ADFL-superHiggs-1}, and Refs. therein).

\begin{table}[h]
\begin{center}
\begin{tabular}{|c||c|}
\hline
$
\begin{array}{c}
\end{array}
$ & $\mathcal{N}=2~$decomposition$~~$ \\ \hline
$
\begin{array}{c}
\\
\mathcal{N}=8 \\
~
\end{array}
$ & $G+6G_{3/2}+15G_{V}+10G_{H}$ \\ \hline
$
\begin{array}{c}
\\
\mathcal{N}=6 \\
~
\end{array}
$ & $G+4G_{3/2}+7G_{V}+4G_{H}$ \\ \hline
$
\begin{array}{c}
\\
\mathcal{N}=5 \\
~
\end{array}
$ & $G+3G_{3/2}+3G_{V}+G_{H}$ \\ \hline
$
\begin{array}{c}
\\
\mathcal{N}=4 \\
~
\end{array}
$ & $G+2G_{3/2}+G_{V}+n\left( G_{V}+G_{H}\right) $ \\ \hline
$
\begin{array}{c}
\\
\mathcal{N}=3 \\
~
\end{array}
$ & $G+G_{3/2}+n\left( G_{V}+G_{H}\right) $ \\ \hline
\end{tabular}
\end{center}
\caption{Decomposition of the massless multiplet content of $\mathcal{N}%
\geqslant 3$, $D=4$ supergravities in terms of $\mathcal{N}=2$ multiplets. $%
n $ denotes the number of matter (vector) multiplets in $\mathcal{N}=3,4$
matter coupled theories. (Massless) gravitino multiplets are not considered.
}
\end{table}

By denoting the contribution of $G$, $G_{3/2}$, $G_{V}$ and $G_{H}$ to the
coefficient $\Delta S_{\mathcal{N}=2}$ (recall (\ref{gen-1})) by $\Delta
S_{2}$, $\Delta S_{3/2}$, $\Delta S_{V}$ and $\Delta S_{H}$ respectively,
the general $\mathcal{N}=4$ result (\ref{N=4-res}) implies the following two
relations:
\begin{eqnarray}
\Delta S_{V} &=&-\Delta S_{H};  \label{N=2-res-1} \\
2\Delta S_{3/2} &=&-\Delta S_{2}-\Delta S_{V}.  \label{N=2-res-2}
\end{eqnarray}
Therefore, by using the decompositions reported in Table 2 as well as the
results (\ref{N=2-self-mirror}) (for self-mirror $\mathcal{N}=2$ $stu$
model), (\ref{N=4-res}) and \cite{Sen-1}
\begin{equation}
\Delta S_{\mathcal{N}=8}=-2,
\end{equation}
one can compute the contribution of each type of massless $\mathcal{N}=2$
multiplet to the total $\Delta S_{\mathcal{N}=2}$ :
\begin{equation}
\Delta S_{2}=\frac{23}{24};~~\Delta S_{3/2}=-\frac{11}{24};~~\Delta S_{V}=-%
\frac{1}{24};~~\Delta S_{H}=\frac{1}{24},  \label{DeltaS-N=2}
\end{equation}
consistent with (\ref{Sen-N=2}) and (\ref{N=2-res-1})-(\ref{N=2-res-2}).
Thus, by exploiting results (\ref{DeltaS-N=2}), Table 2 allows one to
compute the total $\Delta S_{\mathcal{N}}$ for all $\mathcal{N}\geqslant 3$
theories; in particular, the curiously simple result (\ref{main-1}) for $%
\mathcal{N}\geqslant 3$ is obtained.

\section{\label{General-Rels}General Relations between $A_{\mathcal{N}}$, $%
\widetilde{A}_{\mathcal{N}}$ and $\Delta S_{\mathcal{N}}$}

In order to derive Eq. (\ref{main-2}), one just needs to combine the results
(\ref{main-1}) (for $\mathcal{N}\geqslant 3$), (\ref{DeltaS-N=2}) (for $%
\mathcal{N}=2$) with the explicit computation of the coefficient $A_{%
\mathcal{N}}$ of the on-shell gravitational trace anomaly of the fully
undualized theories, whose field content is defined in the $M$-theoretical
setting of \cite{DF-1} (see \textit{e.g.} Table I therein). Nicely, the
following simple and completely general formula is achieved:
\begin{equation}
A_{\mathcal{N}}=2\Delta S_{\mathcal{N}}+\mathcal{N}-4.
\label{rel-AN-DeltaSN}
\end{equation}
Note that, for (generalized) self-mirror theories, Eq. (\ref{rel-AN-DeltaSN}%
) consistently reduces to the result (\ref{main-1}) (made explicit in (\ref
{DeltaS-N-scheme})). Therefore, (\ref{rel-AN-DeltaSN}) is nothing but a
generalization of (\ref{main-1}) for completely generic theories. Note that (%
\ref{rel-AN-DeltaSN}) can actually be extended to include $\mathcal{N}=1$
theories obtained as truncation of $\mathcal{N}=2$ theories, which are
treated in Subsec. \ref{N=1-as-Trunc-N=2}, where the result (\ref{DeltaS-N=1}%
) is derived.

Let us here recall that the coefficients $\widetilde{A}_{\mathcal{N}}$ of
the on-shell gravitational trace anomaly of the fully dualized theories, in
which only physical spin degrees of freedom are present\footnote{%
It should be pointed out that the quantity $K_{0}$ given by Eq. (7.3) of
\cite{Sen-2} is nothing but $\widetilde{A}_{\mathcal{N}}$ itself.}, are
computed in detail in App. A of \cite{DF-1}. By comparing $\widetilde{A}_{%
\mathcal{N}}$ with its undualized counterpart $A_{\mathcal{N}}$, one obtains
the simple and general relation
\begin{equation}
\widetilde{A}_{\mathcal{N}}-A_{\mathcal{N}}=3-\mathcal{N},
\label{rel-AtildeN-AN}
\end{equation}
which, by using (\ref{rel-AN-DeltaSN}), finally yields the general result (%
\ref{main-2}). Note that (\ref{rel-AtildeN-AN}) is independent on the matter
content of $\mathcal{N}\leqslant 4$ theories.

For all (generalized) $\mathcal{N}\geqslant 1$, $D=4$ self-mirror theories,
which all have $A_{\mathcal{N}}=0$ \cite{DF-1}, (\ref{main-2}) reduces to (%
\ref{main-2-self-mirror}). Furthermore, for such theories it also holds
\begin{equation}
\widetilde{A}_{\mathcal{N}}=-\widetilde{A}_{8-\mathcal{N}}-2,
\label{further-rel-self-mirror}
\end{equation}
which is a consequence of the fermion symmetry displayed by Eq. (\ref{main-2}%
). Eq. (\ref{further-rel-self-mirror}) can also be summarized by the
following symmetric pattern, centered at $\mathcal{N}=3$:
\begin{equation}
\begin{array}{cccccccccc}
\mathcal{N}: & 8 & \left( 7\right) & 6 & 5 & 4 & 3 & 2 & 1 & 0 \\
\widetilde{A}_{\mathcal{N}}: & -5 & \left( -4\right) & -3 & -2 & -1 & 0 & 1
& 2 & 3?
\end{array}
,
\end{equation}
providing an hint for a possible ``generalized self-mirror'' $\mathcal{N}=0$%
, $D=4$ gravity theory with $\widetilde{A}_{\mathcal{N}=0}=3$.

\section{\label{N=1-Theories}$\mathcal{N}=1$ Theories with Extremal Black
Holes}

\subsection{\label{N=1-as-Trunc-N=2}$\mathcal{N}=1$ as Truncation of $%
\mathcal{N}=2$}

One can further decompose $\mathcal{N}=2$ massless multiplets $\left\{
G_{\lambda _{\max }}\right\} $ in terms of the $\mathcal{N}=1$ multiplets $%
\left\{ g_{\lambda _{\max }}\right\} $ ($\lambda _{\max }=2,3/2,1,1/2$),
obtaining (see \textit{e.g.} \cite{N=2-red,ADFL-superHiggs-1}, and Refs.
therein)
\begin{eqnarray}
G &=&g+g_{3/2};  \label{N=2-N=1-1} \\
G_{3/2} &=&g_{3/2}+g_{V}; \\
G_{V} &=&g_{V}+g_{c}; \\
G_{H} &=&2g_{c},  \label{N=2-N=1-4}
\end{eqnarray}
where $g\equiv g_{2}$, $g_{3/2}$, $g_{V}\equiv g_{1}$ and $g_{H}\equiv
g_{1/2}$ stand for the graviton, gravitino, vector and chiral massless $%
\mathcal{N}=1$ multiplets, respectively.\medskip

It should be stressed that our treatment of $\mathcal{N}=1$ theories relies
on at least four assumptions:

\begin{enumerate}
\item  in order to display ``large'' extremal BH solutions, $\mathcal{N}=1$
theories should at least contain one vector field : $n_{V}\geqslant 1$ \cite
{ADFT-N=1};

\item  for $n_{c}\geqslant 1$, an attractor dynamics takes place in the
near-horizon geometry \cite{ADFT-N=1};

\item  the results for $\mathcal{N}=1$ theories can be derived from the $%
\mathcal{N}=2$ ones in a purely kinematical way (\textit{i.e.} by multiplet
decomposition). In particular, fermionic bilinear terms coupled to $2$-form
field strengths (see \textit{e.g.} \cite{N=1}) should generally appear for
our analysis to make sense;

\item  the results on $\Delta S_{\mathcal{N}\geqslant 2}$ for ``large''
extremal BPS BHs can be used to derive results on $\Delta S_{\mathcal{N}=1}$
of ``large'' extremal BHs in $\mathcal{N}=1$ theories, in which there is no
central extension of the local supersymmetry algebra, and thus no BPS
notion, at all\footnote{%
Short $\mathcal{N}=2$ BPS massive multiplets are the same as $\mathcal{N}=1$
massive multiplets; for example, a massive hypermultiplet is the same as a
massive charged chiral multiplet \cite{ADFL-superHiggs-1}. Thus, the
multiplet structure of $\mathcal{N}=2$ BPS BHs is the same as $\mathcal{N}=1$
(necessarily non-BPS) BHs \cite{ADFT-N=1}.}.
\end{enumerate}

Also as a consequence of assumptions 1-4, we are therefore assuming that the
kinematical consistent truncation procedure $\mathcal{N}=2\rightarrow
\mathcal{N}=1$ properly takes into account the corresponding change in the
species of bilinear fermionic interaction terms with $2$-form field
strengths (as understood in Secs. \ref{N=2-Multiplet-Decomp} and \ref
{General-Rels}, this issues does not arise for $\mathcal{N}\geqslant 2$%
-extended supergravities, which all have the same Lagrangian structure).

As discussed in \cite{ADFT-N=1}, at least those $\mathcal{N}=1$ theories
obtained as consistent truncations of $\mathcal{N}=2$ ones do satisfy the
conditions of points 1 and 2.\medskip

By using Eqs. (\ref{DeltaS-N=2}) and (\ref{N=2-N=1-1})-(\ref{N=2-N=1-4}),
the contribution of each $\mathcal{N}=1$ multiplet to the coefficient $%
\Delta S_{\mathcal{N}=1}$ (recall (\ref{gen-1})) can be computed; denoting
the contribution of $g$, $g_{3/2}$, $g_{V}$ and $g_{c}$ to $\Delta S_{%
\mathcal{N}=1}$ by $\Delta s_{2}$, $\Delta s_{3/2}$, $\Delta s_{V}$ and $%
\Delta s_{c}$ respectively, one obtains
\begin{equation}
\Delta s_{2}=\frac{65}{48};~~\Delta s_{3/2}=-\frac{19}{48};~~\Delta s_{V}=-%
\frac{3}{48};~~\Delta s_{c}=\frac{1}{48},  \label{DeltaS-N=1}
\end{equation}
thus yielding the general formula:
\begin{equation}
\Delta S_{\mathcal{N}=1}=\frac{1}{48}\left( 65+n_{c}-3n_{V}\right) .
\label{N=1-gen}
\end{equation}
For generalized self-mirror $\mathcal{N}=1$ theories ($n_{c}=3n_{V}+7$) \cite
{DF-1}, Eq. (\ref{N=1-gen}) yields
\begin{equation}
\text{gen.~self-mirror}:\Delta S_{\mathcal{N}=1}=\frac{3}{2},
\label{N=1-gen-self-mirror}
\end{equation}
consistent with the $\mathcal{N}=1$ case of Eq. (\ref{main-1}).

\subsection{\label{N=1-Theories-mc}``Minimally Coupled'' $\mathcal{N}=1$}

On the other hand, (at least) another class of $\mathcal{N}=1$, $D=4$
theories, complementary to the one discussed above, can be considered. Such
a class, which we will dub ``minimally coupled'' (mc), cannot be obtained as
consistent truncation of $\mathcal{N}=2$ theories, and its kinetic vector
matrix is constant: $f_{IJ}\sim \delta _{IJ}$ ($I,J=1,...,n_{V}\geqslant 1$%
). This implies that the complex scalar fields from the $n_{c}$ chiral
multiplets are not involved in an attractor dynamics in the near horizon
geometry of the ``large'' extremal BH under consideration\footnote{%
In fact, they behave as hypermultiplets' scalars in $\mathcal{N}=2$ theories.%
}, which at bosonic level can thus be regarded as a Reissner-N\"{o}rdstrom
(RN) extremal BH coupled to a set of spectator scalar fields and uncharged
vectors.

For ``minimally coupled'' $\mathcal{N}=1$ theories, the contributions to $%
\Delta S_{\mathcal{N}=0}$ split into two parts:

\begin{enumerate}
\item  The $\mathcal{N}=1$ supersymmetrization $\Delta S_{\mathcal{N}=1}^{RN~%
\text{extr}}$ of the RN contribution $\Delta S_{RN~\text{extr}}$, which is
composed by an $\mathcal{N}=1$ graviton multiplet and an $\mathcal{N}=1$
vector multiplet. By making use of Eq. (\ref{DeltaS-N=1}), the resulting
contribution to the logarithmic correction coefficient can be computed to
be:
\begin{equation}
\Delta S_{\mathcal{N}=1}^{RN~\text{extr}}\equiv \Delta s_{2}+\Delta s_{V}=%
\frac{31}{24}.
\end{equation}
This $\mathcal{N}=1$ supersymmetrization of the RN contribution can also be
justified by observing that $\mathcal{N}=3$ ``pure'' supergravity \cite
{FSZ,FSZ-2} displays a ($\frac{1}{3}$-)BPS extremal dyonic RN BH solution,
with entropy \cite{ADF-U-duality-D=4}
\begin{equation}
S_{0}=\frac{\pi }{2}\sum_{i=1}^{3}\left[ \left( p^{i}\right) ^{2}+q_{i}^{2}%
\right] .
\end{equation}
Since there are no scalars, from this system one can derive two $\mathcal{N}%
<3$ Maxwell-Einstein systems, namely $\mathcal{N}=2$ ``pure'' supergravity
\cite{N=2-pure} and $\mathcal{N}=1$ supergravity coupled to $1$ vector
multiplet \cite{N=1-nV=1}, the two theories being related by exchanging one
gravitino with one gaugino (with related interactions; see also \textit{e.g.}
the discussion in \cite{ADFT-N=1}).

\item  The $\mathcal{N}=1$ supersymmetrization $\Delta S_{\mathcal{N}%
=1}^{GB} $ of the Gauss-Bonnet (GB) contribution $\Delta S^{GB}$, which is a
well-defined, independent invariant in $\mathcal{N}=1$, $D=4$ superspace
\cite{FZ-N=1}. We start from the non-supersymmetric expression (see \textit{%
e.g.} \cite{Sen-2}, and Refs. therein)
\begin{equation}
\Delta S^{GB}=-\frac{1}{360}\left( n_{s}+62n_{V}+\frac{11}{2}n_{MF}\right) ,
\label{GB}
\end{equation}
where $n_{s}$, $n_{V}$ and $n_{MF}$ respectively denote the number of real
scalar, vector and $\lambda =1/2$ Majorana spinor massless fields. By
recalling the helicity content of $\mathcal{N}=1$ massless multiplets, it is
immediate to re-express the right-hand side of (\ref{GB}) in terms of only $%
n_{c}$ and $n_{V}$, where the latter now stands for the number of $\mathcal{N%
}=1$ massless vector multiplets other than the one contained in the $%
\mathcal{N}=1$ supersymmetrization of RN term. Thus, one obtains the
consistent $\mathcal{N}=1$ supersymmetrization of $\Delta S^{GB}$:
\begin{equation}
\Delta S_{\mathcal{N}=1}^{GB}=-\frac{1}{360}\left( 6n_{c}+\frac{135}{2}%
n_{V}\right) .
\end{equation}
\end{enumerate}

Summing all up, in the class of $\mathcal{N}=1$ theories under
consideration, the total contribution to the logarithmic correction
coefficient reads as follows ($n_{V}\geqslant 0$):
\begin{equation}
\Delta S_{\mathcal{N}=1}^{\text{mc }}=\Delta S_{\mathcal{N}=1}^{RN~\text{extr%
}}+\Delta S_{\mathcal{N}=1}^{GB}=-\frac{1}{360}\left( -465+6n_{c}+\frac{135}{%
2}n_{V}\right) ,  \label{DeltaS-N=1-mc}
\end{equation}
which is completely different from Eq. (\ref{N=1-gen}).

Note that in ``minimally coupled'' $\mathcal{N}=1$ supergravities the vector
multiplets participating in the $\mathcal{N}=1$ supersymmetrization of the
RN term stands on a different footing with respect to the other $n_{V}$
vector multiplets\footnote{%
The Noether supercurrent coupling to the gravitino \cite{N=1-nV=1} $%
J_{\alpha }^{\mu }\psi _{\mu }^{\alpha }$ introduces a fermionic bilinear
term proportional to the flux of the RN vector field (see also \cite{N=1}).
Note that only the gaugino of the $\mathcal{N}=1$ vector multiplet of the $%
\mathcal{N}=1$ RN term interacts with the gravitino; indeed, all other
vector fields are minimally coupled, and they have vanishing fluxes of the
corresponding $2$-form field strengths. This explains why the $J\psi \sim $Re%
$\left( \overline{\lambda }\psi F\right) $ interactions do not contribute to
the $\mathcal{N}=1$ supersymmetrization of the GB term (discussed at point 2
above).}. Furthermore, by construction, the $\mathcal{N}=0$ limit of the
expression (\ref{DeltaS-N=1-mc}) corresponds to the $\Delta S_{\mathcal{N}%
=0} $ given by Eq. (1.3) of \cite{Sen-2}; indeed, when $f_{IJ}\sim \delta
_{IJ}$, by setting $\psi =0$ all bilinear fermionic terms coupled to the $2$%
-form vector field strengths vanish. In particular, the RN limit of (\ref
{DeltaS-N=1-mc}), which amounts to setting $n_{c}=n_{V}=0$ and to removing
the gravitino and gaugino contained in the $\mathcal{N}=1$ supersymmetric RN
term, correctly yields the non-supersymmetric RN contribution \cite{Sen-2}
\begin{equation}
\Delta S^{RN~\text{extr}}=-\frac{241}{90}.
\end{equation}
\smallskip

Finally, let us shortly comment on the physical significance of our results.

For generic theories, the result (\ref{main-2}) expresses the fact that the
entropy correction is the same as the (on shell) gravitational anomaly, up
to an universal shift. On the other hand, for (generalized) self-mirror
theories, Eq. (\ref{main-1}) implies that $\Delta S_{\mathcal{N}}$ is odd
under the fermionic symmetry $\mathcal{N}\rightarrow 8-\mathcal{N}$, and
that it is given by the lowest helicity component of the gravity multiplet,
namely $\lambda _{\min }=2-\mathcal{N}/2$. Thus, Eqs. (\ref{gen-1}) and (\ref
{main-1}) yield the following correction to the Bekenstein-Hawking
entropy-area formula of (generalized) self-mirror theories:
\begin{equation}
S=\frac{A_{H}}{4}+\left( 2-\mathcal{N}/2\right) \ln \left( A_{H}^{2}\right) .
\end{equation}
This is universal, because it only depends on $\mathcal{N}$, and it
increases or decreases the classical Bekenstein-Hawking entropy depending on
$\mathcal{N}<4$ or $\mathcal{N}>4$ (it is vanishing for $\mathcal{N}=4$).

\section*{Acknowledgments}

We would like to thank Ashoke Sen for enlightening correspondence.

The work of S. F. is supported by the ERC Advanced Grant no. 226455, \textit{%
``Supersymmetry, Quantum Gravity and Gauge Fields''} (\textit{SUPERFIELDS}),
and in part by DOE Grant DE-FG03-91ER40662.

\end{document}